\documentclass{aa}  
\usepackage{graphicx}
\usepackage{txfonts}
\usepackage{natbib}
\bibpunct{(}{)}{;}{a}{}{,}
\pdfoutput=1 
\newcommand{\doceCO}{\mbox{$^{12}$CO}}
\newcommand{\goha}{\mbox{Gomez's Hamburger}}
\newcommand{\gohab}{\mbox{GoHam}}
\newcommand{\goham}{\mbox{Gomez's Hamburger}}
\newcommand{\treceCO}{\mbox{$^{13}$CO}}
\newcommand{\jdu}{\mbox{$J$=2$-$1}}
\newcommand{\juc}{\mbox{$J$=1$-$0}}

\newcommand{\kms}{\mbox{km\,s$^{-1}$}}

\newcommand{\Jybeam}{\mbox{Jy\,beam$^{-1}$}}
\newcommand{\ms}{\mbox{$M_{\mbox{\sun}}$}}
\newcommand{\ls}{\mbox{$L_{\mbox{\sun}}$}}

\newcommand{\TB}{\mbox{$T_{\mathrm B}$}}

\newcommand{\lsim}{\raisebox{-.4ex}{$\stackrel{\sf <}{\scriptstyle\sf \sim}$}}

\newcommand{\farcss}{\mbox{\rlap{.}$''$}}
\begin{document}
   \title{Gomez's Hamburger (IRAS\,18059-3211): 
A pre main-sequence A-type star 
}

   \author{
          V. Bujarrabal\inst{1}
	  \and
	K. Young\inst{2}
          \and
          D. Fong\inst{3}
          }

   \offprints{V. Bujarrabal}

   \institute{             Observatorio Astron\'omico Nacional (OAN-IGN),
              Apartado 112, E-28803 Alcal\'a de Henares, Spain\\
              \email{v.bujarrabal@oan.es}
              \and
 Harvard-Smithsonian Center for Astrophysics, 60 Garden Street,
              Cambridge, MA 02138, USA                 \\
              \email{rtm@cfa.harvard.edu}
\and 
 Submillimeter Array, Harvard-Smithsonian Center for Astrophysics, 645
North A'ohoku Place, Hilo, HI 96720, USA 
\\
\email{dfong@sma.hawaii.edu  } 
          }

   \date{accepted for publication in A\&A}

  \abstract
   {}
   {We study the nature of Gomez's Hamburger
  (IRAS\,18059-3211; \gohab), a nebula that has been proposed to be
  a post-AGB object. Such a classification is not confirmed; instead, we
  argue that it will be a key object in the study of disks rotating
  around young stars.}
   {We present high resolution SMA maps of CO $J$=2--1 in Gomez's
  Hamburger. The data are analyzed by means of a code that simulates
  the emission of a nebula showing a variety of physical conditions and
  kinematics.}
   {Our observations clearly show that the CO emitting gas in Gomez's
  Hamburger forms a spectacular disk in keplerian rotation. Model 
  calculations undoubtly confirm this result. The central (mainly 
  stellar) mass is found to be high, $\sim$ 4 \ms\ for a distance of 500
  pc. The mass and (relatively low) luminosity of the source are,
  independent of the assumed distance, very different from those 
  possible in evolved stars. This object is probably 
  transitional between the pre-MS and MS phases, still showing
  interstellar material around the central star or stellar system. 
}  
   {}
   \maketitle
%

\section{Introduction}

Gomez's Hamburger (IRAS\,18059-3211; hereafter \gohab) is a southern
nebula with a very spectacular appearance in the optical
\citep[see][and HST images in press release number
2002/19]{ruizetal87}: an obscuring lane separates two flat, bright
regions, presumably illuminated by a central star that remains hidden
by the equatorial disk. The nebula was originally identified on a plate
by A.\ G\'omez, in 1985. The central star has been classified as an
A0-type star, from spectroscopic analysis of the scattered light
\citep{ruizetal87}. Its spectral energy distribution (SED) shows two
maxima, in the optical and FIR
\citep{ruizetal87}. Some emission excess at 2-10 $\mu$m also appears,
probably due to relatively hot dust.  Similar features are often
detected in pre-MS and post-AGB sources, they have been proposed to be
due to reservoirs of material relatively close to the star,
i.e.\ rotating disks
\citep[e.g.][]{chianggoldr97,wood02,vanwinckel03}. In our case, the
relatively low intensity of the optical maximum is due to strong
absorption by grains in the disk, seen almost edge-on, the FIR maxima
being due to reprocessed light.

\citet{ruizetal87} proposed that \gohab\ is a post-AGB nebula, mainly
based on similarities with some of such objects, like the Red
Rectangle and M\,1--92, and the absence of clear association with
interstellar clouds. We also note the absence of conspicuous emission
components in the Balmer lines, which are typical of protostars.

We present CO $J$=2--1 high-resolution mapping of \gohab\ that
undoubtly indicates that the nebula is essentially a disk in keplerian
rotation. As we will see, the rotation velocity yields a relatively
high mass for the central star(s). We will argue that the central
stellar mass and luminosity are not compatible with those acceptable
for post-AGB sources, but are very similar to those expected in pre-MS
A-type stars.

\section{Observations}

The observations were obtained with the Submillimeter Array (SMA) on
the night of 2006 June 6 (UT), with the array in the extended
configuration.  All eight antennas were operational.  The J2000
coordinates used for IRAS\,18059-3211 were R.A.\ = 18:09:13.37, Dec =
--32:10:49.5. The receivers were tuned to place the \doceCO\ $J$=2--1
transition at 230.53797 GHz in the upper sideband, and the \treceCO\
2--1 transition at 220.39868 GHz in the lower sideband.  During the
track the precipitable water vapor above the array ranged from 1.2 to
2.0 mm.  The DSB receiver temperatures ranged from 100\,--$>$140 K near
source transit, to 160\,--$>$190 K at the lower elevation limit of 15
degrees.  The correlator was configured to give a spectral resolution
of 203.125 kHz ($\sim$0.26 \kms) per channel in the regions of the IF
containing the CO lines, and a coarser 3.25 MHz ($\sim$4.23 \kms) per
channel in the remaining regions of the full 2 GHz IF bandwidth.
Ganymede was used as a flux calibrator, 3C279 was the bandpass
calibrator, and observations of the quasars 1911-201 (2.3 Jy) and
1924-292 (4.4 Jy) were made every 20 minutes, to track changes in the
instrument's complex gain.  IRAS 18059-3211 was observed from 08:45 to
15:00 UT, and 4.6 hours of on-source data was acquired.  Projected
baseline lengths ranged from 12 to 140 k$\lambda$.

   The MIR package was used for calibration.  Antenna-based bandpass
spectra were derived from the crosscorrelation data, and applied to
flatten the spectral response.  Next, pseudo-continuum data were
derived by vector averaging the spectral channels in the region of the
bandpass free of line emission.  This pseudo-continuum channel was used
to derive the fluxes of the quasars.  The interleaved quasar
observations were then used to calibrate the IRAS 18059-3211 data.  The
data were then exported to the AIPS package for mapping.  Continuum
emission was clearly detected in our source, and it was subtracted from
the channels showing line emission before mapping (see Fig.\ 1). 
The measured continuum flux at 1.3 mm is $\sim$ 0.3 Jy; its deconvolved
image is about 1$''$ wide and elongated in the direction of the
nebula's major axis, at a position angle of about --5 degrees (measured
from north to east). The J2000 coordinates of the continuum source
center are R.A.\ = 18:09:13.42, Dec = --32:10:50.0.

From checks on the positions of the calibrators, we estimate that the
accuracy in the absolute coordinates is $\sim$ 0\farcss3.

We present here the $^{12}$CO $J$=2--1 maps, Figs.\ 1 and 2. The
1$\sigma$ noise in the channel maps is 0.1 \Jybeam. The
half-maximum size of the clean beam is 1.47x1.09 arcsec, the P.A.\ of
the major axis being --10 degrees. The corresponding conversion factor
from flux units to Rayleigh-Jeans--equivalent brightness temperature is
14.3 K/(Jy/beam). In Fig.\ 3 we show the line profiles of \doceCO\ and
\treceCO\ \jdu\ resulting from integration over the emitting
region. The $^{13}$CO $J$=2--1 observations, as well as $^{12}$CO 3--2
and 6--5 data obtained in a recent observing run, are still under
analysis.

We also show in Fig.\ 1, first panel, and image obtained from the
HST archive. This image was taken on April 12, 2006; HST project 10603,
P.I.: D.\ Padgett. The NICMOS camera was used, with the F110W filter
(wide filter centered on 1.10 microns) and an exposure time of 768
seconds. The position and orientation of the HST image is compatible
with our mm-wave maps within the astrometric uncertainties.

   \begin{figure*}
\vspace{-2.0cm}
   \hspace{-2.2cm}
\rotatebox{270}{\resizebox{18.6cm}{!}{ 
\includegraphics{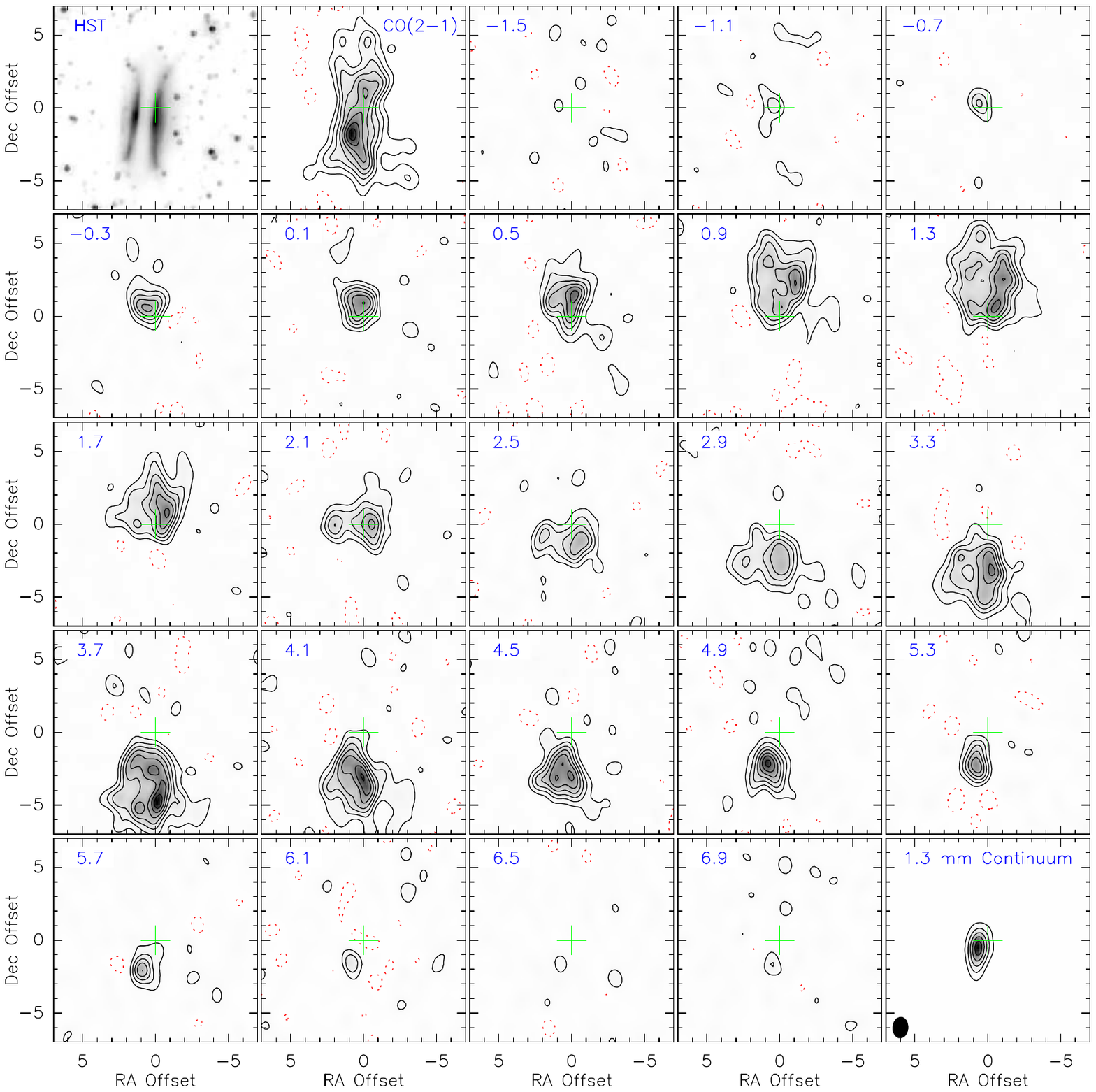}
}}
\vspace{-1.0cm}
   \caption{Channel maps of the \doceCO\ \jdu\ line from Gomez's
   Hamburger, continuum has been subtracted. First contour and contour
   step are 0.2 Jy/beam; negative values are represented by dashed
   contours. The LSR velocity in \kms\ for the center of each channel
   is indicated in the upper left corner. The J2000 coordinates for the
   reference position, the cross in the maps, are R.A.\ = 18:09:13.37,
   Dec = --32:10:49.5. The first panel shows an HST/NICMOS image
   of this source.  In the second panel we represent the map of the
   velocity-averaged CO intensity (first contour and step: 0.05
   Jy/beam). The last panel gives the continuum intensity distribution
   (first contour and step: 0.03 Jy/beam) and the clean beam
   half-intensity size (black ellipse).}
              \label{maps}%
    \end{figure*}

\section{CO emission model}

We have analyzed our data using a model that simulates the emission of
the nebula for a variety of physical conditions and kinematics; it was
soon found that \goham\ is well modelled by a disk in keplerian
rotation whose width increases with the distance to the center. The
code and general features of the model are very similar to those
described by \citet[][]{bujetal03,bujetal05}.

In the line emission model, the populations of the CO rotational levels
are given by a single excitation temperature, which is assumed to be
equal to the kinetic temperature. As discussed in our previous works,
these approximations are valid for the typical densities actually found
from our model fitting ($>$ 10$^5$ cm$^{-3}$) and the high opacities
expected for the $^{12}$CO lines (from comparison of the data presented
here with our preliminary \treceCO\ maps, see Fig.\ 3). We assume a
shape of the CO cloud and a spatial distribution of the velocity,
temperature, density, and CO relative abundance. Given these
parameters, our code calculates the brightness for a number of lines of
sight solving the full radiative transfer equation. Opacity effects and
velocity shifts are accurately taken into account. Such a brightness
distribution is convolved with the cleaned telescope beam, and images
with the same units as the observed ones are produced.

The continuum emission at $\lambda$ = 1.3 mm is treated following
\citet[][]{bujetal05}; note that its contribution is not subtracted in
the observed velocity-position diagram, Fig.\ 2. We assume the detected
mm-wave continuum to be due to optically thin dust emission, since the
measured intensities are compatible with the extrapolation of the FIR
dust emission \citep[][]{ruizetal87}. The free-free emission in young
A-type stars \citep{skinner93} is usually much weaker than the level
detected in our data; in the Red Rectangle, a post-AGB source that
could have similar properties as \gohab, the free-free contribution
to the continuum at this frequency is also much lower than that of dust
emission \citep[e.g.][]{menshetal02}.  In any case, the treatment of
the continuum is the same if there is some significant contribution
from optically thin free-free emission. Note that the very
probably low optical depth of the dust continuum at this frequency
simplifies the treatment and avoids discussions on the dust
characteristic temperature. The spatial distribution of the dust
emission coefficient is described in the code by a three-dimension
gaussian function around the nebula center; the corresponding
parameters are estimated from normalization of the resulting total flux
and extent with respect to the data (Sect.\ 2). The effects of the dust
continuum emission are however very small and hardly visible in the
line maps.

\begin{table*}[bthp]
\caption{Structure and physical conditions in the molecular disk in
  Gomez's Hamburger, derived from our model fitting of the $^{12}$CO
  $J$=2--1 data.  Dependence on the assumed distance is given in the
  relevant cases.  Other parameters of the modeling are also given.}
\begin{center}                                          
\begin{tabular}{|l|lc|l|}
\hline\hline
& & & \\
{\bf Parameter}  & {\bf Law} & {\bf Values} & comments \\ 
& & & \\
\hline\hline
   &  &  &  \\
Outer radius  &   &  $R_{\rm out}$ = 4.75 10$^{16}$ ($\frac{D({\rm pc})}{500}$)
 cm &   \\
   &  &  & \\
\hline\hline
   &  &  &  \\
Inner radius  &  &  $R_{\rm in}$ = 3.5 10$^{15}$ ($\frac{D({\rm pc})}{500}$)
 cm &   \\
   &  &  & \\
\hline
   &  &  &  \\
Disk & linear & $H(R_{out})$ = 2.3 10$^{16}$ ($\frac{D({\rm pc})}{500}$) cm &
   \\
thickness & &  $H(R_{in})$ = 5.4 10$^{15}$ ($\frac{D({\rm pc})}{500}$) cm & \\
   &  &  &  \\
\hline
   &  &  & \\
Tangential & $V_{\rm t} \propto 1/\sqrt{r}$ & $V_{\rm t}(R_{\rm out}/2)$
= 1.6 \kms & \\
velocity  & (keplerian)~~~ & central mass: 4 ($\frac{D({\rm
 pc})}{500}$) \ms &  \\
   &  &  &  \\
\hline
   &  &  &  \\
 Temperature  & $T \propto 1/r^{\alpha_T}$ & $T(R_{\rm out}/2)$ = 21 K &
 plus increase in disk edges  \\ 
 & & $\alpha_T$ = 0.4 & (Sect.\ 3) \\
   &  &  &  \\
\hline
  &  &  &  \\
 Gas density & $n \propto 1/r^{\alpha_n}$ & $n(R_{\rm out}/2)$ = 2.5 10$^6$ 
($\frac{500}{D({\rm pc})}$) cm$^{-3}$ &  plus decrease in outer edges   \\
  &  & $\alpha_n$ = 1 &  (Sect.\ 3) \\
   &  &  &  \\
\hline
\end{tabular}
\begin{tabular}{|l|lc|l|}
\hline
 & & & \\
{\bf Other parameters}  & {\bf Law} & {\bf Values}
 & comments \\ 
& & & \\
\hline\hline
 & & & \\
Axis inclination from the plane of the sky & & 5$^\circ$ & from optical and
CO data  \\
 & & & \\
\hline
 & & & \\
Axis inclination in the plane of the sky (PA) & & 85$^\circ$ & from
optical and CO data \\
 & & & \\
\hline
& & & \\
Distance  & & 500 pc & from possible values of the luminosity and mass  (Sects.\ 4 and 5) \\
 & & & \\
\hline
 & & & \\
CO relative abundance & constant~~~ & ~~10$^{-4}$~~  & from various CO data
 (Sect.\ 3) \\
 & & & \\
\hline
 & & & \\
LSR systemic velocity &  & 2.7 \kms & from CO data  \\
 & & & \\
\hline\hline
\end{tabular}
\end{center}
\end{table*}

The disk shape is defined by its
inner and outer radii ($R_{\rm in}$, $R_{\rm out}$) and thickness ($H$)
at these radii.

The macroscopic velocity is simply a keplerian rotation, defined by the
tangential velocity at a given distance; the data are not compatible
with the presence of significant expansion or infall velocities.  The
local velocity dispersion is assumed to be composed of the thermal
dispersion (given by the kinetic temperature distribution, see below)
and turbulent movements. However, including turbulence was not found to
be necessary in our case to fit the data, whose low velocity dispersion
indicates turbulent velocities \lsim\ 0.1 \kms.

The total density $n$ and temperature $T$ of the disk gas are
assumed to vary with potential laws. Thus, in our model: $n(r)$ = 
$n(R_{\circ}) (R_{\circ}/r)^{\alpha_n}$ and $T(r)$  = 
$T(R_{\circ}) (R_{\circ}/r)^{\alpha_T}$,   
where $n(R_{\circ})$, $T(R_{\circ})$, $\alpha_n$, and $\alpha_T$ are free
parameters. 

We assume a CO relative abundance $X$(CO) = 10$^{-4}$, usually found in
interstellar condensations, but somewhat smaller than the values
typical of young PNe. Note that, due to the assumption of a thermalized
level population, changes in the abundance and density are compensated
for exactly in the model predictions when the product $n X$ remains
constant.

To account for the increase in intensity observed to come from the
edges of the disk (Fig.\ 1), we also included possible variations of
the density and temperature in them. Since the \doceCO\ lines are
optically thick and this limb brightening does not appear in our
\treceCO\ data, such a phenomenon must be mostly due to a higher
temperature and a lower density in the edges (which is in any case
expected in this kind of disk). We have successfully fitted the data
assuming that the outer 20\% of the cone (in the axial direction) shows
a temperature increase by a factor 2 and a decrease in density by a
factor 10 (with respect to the general potential laws).  We note that
the brightness increase is less remarkable in the east part of the disk
than towards the west due to significant selfabsorption.  Some limb
brightening is also found in the outer edge of the cone (in the disk
plane), then implying variations of the same kind in these regions: our
best fit includes an increase of the temperature, by 20\%, in the outer
very diffuse part of the disk (outer 25\%), in which the density
already shows a decrease by a factor of 10. We also assumed an increase
of the temperature of 50\% in the innermost regions, within $R_{\rm
in}$ and $2 R_{\rm in}$. Although we are not comparing here the
physical conditions we find from our simplified fitting with disk
models (to be done in a forthcoming paper), we note that such an
increase of the temperature in disk edges is expected from theoretical
results \citep[e.g.][]{dullemond07}.

Note that the observed disk emission is not exactly symmetrical in
velocity and position with respect to the rotation axis (P.A.\ $\sim$
85$^\circ$; see Figs.\ 1, 2); our model assumes axial symmetry and only
can reproduce an average of both parts. The emission from the outer,
very diffuse edge of the disk (spatial offsets $\sim$ $\pm$ 5$''$,
shift relative to systemic velocity $\sim$ $\pm$ 1.2 \kms, see Fig.\
2), 
in particular, presents a noticeable asymmetry
between the north and south parts of the disk. The asymmetry found in
our maps with respect to disk equator (P.A.\ $\sim$ --5$^\circ$) is due
to the slight inclination of the nebula axis with respect to the plane
of the sky and well reproduced by our model.

The fitted values of the parameters describing the structure,
kinematics and physical conditions in the model nebula are given in
Table 1. 
The parameter values are scaled with the distance; we take $D$
= 500 pc as standard value following our final conclusions on the
possible values of the stellar mass and total luminosity (Sect.\ 5).
The density ranges between 1.5 10$^7$ cm$^{-3}$
(close to the central star) and 1.5 10$^6$ cm$^{-3}$, in the outer
regions (ten times smaller in the very diffuse outermost edge). The
temperature ranges between 65 K (close to the star) and 20 K (in the
coolest equatorial regions).

We note that some of the fitted parameters are almost directly given by
the observational features, their values are therefore quite accurate
and not model dependent. The model fitting of course confirms the
values that we directly measure from the maps. The velocity variation
with the distance to the star is obviously kepler-like and the
measurement of the central mass is accurate, see Fig.\ 2 and further
discussion in Sect.\ 4. The diameter of the disk is also given by the
observations, about $\sim$ 13$''$, equivalent to $\sim$ 10$^{17}$ cm at
$D$ = 500 pc. The total width of the disk (at least its maximum value,
at the end of the disk) can also be measured from our maps, we find a
maximum width $\sim$ 3$''$, 2.3 10$^{16}$ cm at 500 pc. The data also
indicate that the disk width increases more or less linearly with the
distance to the center, the result being a flaring or quasi-flaring
disk. Although such a thick disk cannot be considered to be exactly
'keplerian', the variation of the tangential velocity does follow the
$\frac{1}{\sqrt{r}}$ law. 

Since the \doceCO\ \jdu\ line is found to be optically thick, the
dependence of its intensity on the density and CO abundance is small, and
these parameters cannot be well determined from our data. The
values used here have been chosen to be compatible with our preliminary
\treceCO\ maps, though no accurate fitting of our observations of both
lines is attempted in this paper. In particular, we cannot give
here an accurate measurement of the disk total mass. 
On the other hand, the fact
that the line is optically thick allows us to determine the
temperature that, regardless of selfabsorption effects, must be similar
to the measured brightness temperature.  

The inner radius of the CO-emitting disk is also not well determined
from our data. We detect CO emission up to $\pm$ 3.5 \kms\ from the
systemic velocity. For the assumed velocity law, this means that we
detect emission from regions farther than $\sim$ 0\farcs5. This value
is similar to the beam radius, as expected, because much smaller
regions are strongly diluted within our spatial resolution and more
difficult to detect. On the other hand, we do not see a clear central
minimum in our maps, confirming that the inner disk radius cannot be
larger than the beam. Regions closer to the center than $\sim$ 0\farcs5
may also be molecule rich, but they are not well probed by our data.

Our synthetic maps essentially depend on the nebula structure, velocity
field and temperature distribution. In this work, we focus on the
determination of the rotation velocity and the central (mainly stellar)
mass, which involves in fact only one free parameter, $V_{\rm t}$ at a
given point. The distance is widely discussed in Sects.\ 4 and 5.
Accordingly, the model remains very simple. The determination of the
physical conditions (density, temperature, CO abundance) will be
improved in a forthcoming paper, from fitting of several
rotational transitions of \doceCO\ and \treceCO\ and a detailed
discussion on the physical meaning of the results and their
compatibility with theoretical considerations.

\section{The mass and luminosity of Gomez's Hamburger}

Our observations clearly show that the CO emitting region in
\goham\ is a disk, strikingly coincident with the equatorial dark lane
and bright rims seen in the optical. The CO-rich disk is even slightly
larger than the optical image and very probably represents the total
nebular component of this source (Fig.\ 1). The position-velocity
diagram along the equatorial direction clearly shows the butterfly-like
shape characteristic of keplerian rotation (Fig.\ 2). Our model of
emission from a rotating disk reproduces very closely the CO line
observations of \gohab, see Fig.\ 2, confirming that the Gomez's
Hamburger nebula is a disk in keplerian rotation.

The optical image and the double-maximum SED of \gohab\ \citep[Sect.\
1,][]{ruizetal87} strongly support the identification of a disk seen
practically edge-on. The energy radiated by our source shows a double
maximum in the optical and FIR, with also some excess at about 2-10
$\mu$m, which is characteristic of a disk rotating around a relatively
hot star. Both young and post-AGB stars surrounded by disks of
dusty material show SEDs of this kind
\citep[e.g.][]{vanwinckel03,wood02}, in which the short-$\lambda$
maximum comes from the stellar emission, absorbed and/or scattered by
the grains, while the FIR maximum is due to the emission of the disk dust,
heated by the stellar radiation.

Our model fits the rotating disk structure for a central mass
$M_s$(\ms) $\sim$ 8 \ms, for a distance $D$ = 1 kpc; $M_s$ varies
proportionally to the assumed distance. It is important to note that
this result is very model-independent. Since the disk is seen almost
edge-on, corrections to the tangential velocity and distance to the
center due to the geometry are negligible. We can directly see from our
maps and velocity-position diagram that gas at a distance of 6
arcseconds (equal to 9 10$^{16}$ cm if $D$ = 1 kpc) rotates at 1.1 \kms.
Errors to these figures, directly taken from the maps, are smaller than
$\sim$ 10\%, including the correction due to projection effects (the
correction to the observed velocity for an inclination of 10$^\circ$,
see Sect.\ 3, is of less than 2\%).  These values of the velocity and
distance also yield, assuming keplerian rotation, a central stellar
mass $M_s$(\ms) = $8\ms \times D$(kpc). The choice of other
velocity/position pairs yield the same mass, because, as previously
mentioned, the velocity field is very exactly keplerian.  Such a mass
is very high for a post-AGB star, since the central stars of planetary
nebulae are expected to have masses smaller than 1 \ms, see e.g.\
\citet{pottasch84}, \citet{blocker95}.

On the other hand, the total luminosity of \gohab,
integrating all wavelengths, is moderate, $\sim$ 35 \ls\ for $D$ = 1
kpc, since the apparent bolometric luminosity is 1.1 10$^{-9}$
erg\,cm$^{-2}$\,s$^{-1}$ \citep[from careful study of the SED
by][]{ruizetal87}. \citet{ruizetal87} compared the (apparent) SED of
\gohab\ with those of two well known post-AGB objects, the
Red Rectangle and M\,1--92 (see their Fig.\ 4). \gohab\ is
more than ten times weaker at all wavelengths than M\,1--92 and about
100 times weaker than the Red Rectangle. M\,1--92 has an absolute
luminosity of 10$^4$ \ls\ and a distance of about 2.5 kpc; the Red
Rectangle has a luminosity $\sim$ 3.5 10$^3$ \ls, assuming a distance
of 710 pc \citep[e.g.][]{bujetal01,menshetal02}. Both are typical
post-AGB stars, and other objects in this evolutionary phase show very
similar absolute luminosities. We also note that the mass of the binary
system in the center of the Red Rectangle is well known, 1.5 \ms,
measured from the (almost) keplerian disk around it.

Accordingly, for \gohab\ to have a post-AGB luminosity, at
least, $\sim$ 3 10$^3$ \ls, it should be placed at a surprising distance
$D$ $\sim$ 10 kpc. This would imply an exceptional central mass $M_s$
$\sim$ 80 \ms, as well as a distance to the galactic plane $\sim$ 
1 kpc, particularly unexpected for such a massive object. Even assuming
that there is a binary in the center, we should need something like two
40\ms\ stars. 

If we assume that the stellar component is a binary and most of the
mass is in the less evolved star, it should have a mass of almost 80
\ms. Then, the evolved star should have had an initial mass
still larger (since it evolved faster), leading to still more
surprising conclusions. 

These conditions can be somewhat relaxed if we take into account that
the stellar short-$\lambda$ emission is absorbed by dust before it
leaves the nebula. \citet{ruizetal87} suggest a reddening $E(B-V)$
$\sim$ 0.56; we must take into account that the interstellar extinction
measured in the direction of our source is $\sim$ 0.5 mag
\citep[][]{dobashi05}. In any case we note that a significant
absorption in the optical in the direction in the equatorial plane is
clearly suggested by the nebula image. This effect would imply a
strongly anisotropic emission in the optical and UV, being minimum in
the direction of the equator (i.e.\ toward us). We can easily check
that the correction due to the above extinction value is of the order
of a factor ten (at short $\lambda$). Larger correction values cannot
be obtained for reasonable geometrical distributions of the
circumstellar dust. Taking into account that the total emission from
the stellar component of the actually observed SED is about 10 times
smaller than the dust emission component, we conclude that the
correction to the total luminosity due to extinction would increase it
by a about factor two, up to 70 \ls\ at $D$ = 1 kpc. The
'post-AGB distance' estimated for this value of the luminosity would be
$\sim$ 7 kpc, and the central mass should be close to 60 \ms, still far
too large. 
The correction of the extinction effects is uncertain, but in fact
can hardly be much larger than this value, since the fraction of the
total stellar light absorbed by the dust lane and reemitted in the FIR
cannot be negligible. For the disk aperture deduced from our model
fitting, about 1/3 of the total stellar light is intercepted by the
dust lane. Therefore, the total luminosity would be at most three times
larger than the FIR luminosity (about 100 \ls, at 1 kpc), corresponding
to the extreme case of negligible contribution of direct and scattered
stellar light to the detected fluxes. Even in this case, the effects on
the mass are very moderate, and a too large central stellar mass of
$\sim$ 45 \ms\ would be required.  A similar correction is found in the
models developed by \citet{menshetal02} for the Red Rectangle, a
post-AGB star showing a massive rotating disk that is seen almost
edge-on. We note that in other post-AGB stars showing rotating disks,
the extinction due to dust is always moderate and the effects of the
viewing angle on the luminosity determination are expected to be almost
negligible \citep[e.g.][]{vanwinckel03}.

Even if we are now checking if \gohab\ can be a post-AGB object,
we will also consider the effects of the viewing angle on the
luminosity determination predicted by models of disks around
pre-main-sequence stars. Moderate effects not exceeding changes by a
factor two in the total luminosity are found by fitting of the SED 
using the models for young stellar objects by \citet[][]{robi07}.
\cite{wood02} published a grid of calculations for different disk
models. In extreme cases, these calculations yield a correction to the
total luminosity of about a factor 10. Even adopting such a large
correction, we would derive a too high value of the central stellar
mass, $\sim$ 25 \ms. We recall that these models are not well adapted
to our object, mainly if we assume by the moment that it is in the
post-AGB phase. Finally we note that the reasons why even very large
corrections to the luminosity estimate cannot yield mass values
comparable to that of an evolved star are: a)
the luminosity depends on $D^2$, while the mass derived from the
velocity pattern varies like $D$, so large variations of the luminosity
yield much smaller variations of the mass, and b) the mass of post-AGB
stars are expected to be significantly smaller than 1 \ms, typically
$\sim$ 0.5 \ms, really very far from the values we derive.

Another option is to assume that the central mass is not stellar, but
corresponds to a very massive and compact nebular component. Young
planetary nebulae are much less massive, at most $\sim$ 1 \ms, and not
so compact \citep[see][]{bujetal01}. There are two yellow hypergiants,
IRC+10420 and AFGL\,2343, surrounded by nebulae with about 5 \ms\
\citep{ccarrizo07}, but these nebulae are spherical and very extended,
both objects are placed at less than 500 pc from the galactic plane,
and both have luminosities $\sim$ 5 10$^5$ \ls.

We note, moreover, that at a distance of $\sim$ 10 kpc, the 
rotating disk would be a monster, with a total radius of $\sim$
10$^{18}$ cm and a total nebular mass larger than 10 \ms, nothing
comparable to known disks rotating around post-AGB stars
\citep[see][]{bujetal05,bujetal07,vanwinckel03}.

   \begin{figure}
\vspace{-0.3cm}
   \hspace{-0.1cm}
\rotatebox{0}{\resizebox{9cm}{!}{ 
\includegraphics{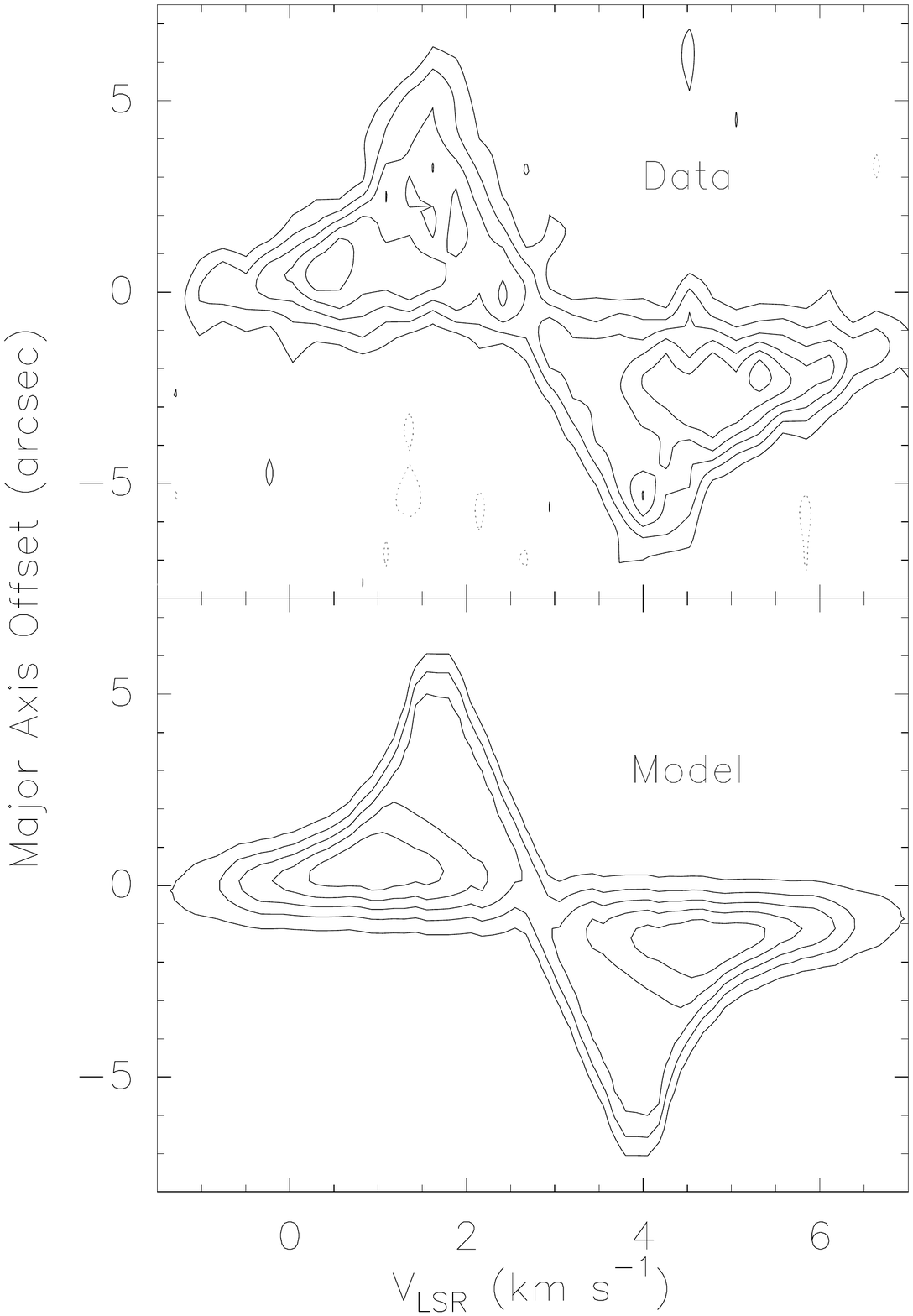}
}}
\vspace{-0.8cm}
   \caption{Velocity-position diagram of the \doceCO\ \jdu\ line
   emission from Gomez's Hamburger (upper panel) along a direction
   defined by P.A.: --5$^\circ$. Lower
   panel: model fitting of the observed velocity-position diagram. In
   both cases, the same contours as in Fig.\ 1 are used, but continuum
   is not subtracted in these diagrams.}
              \label{maps}%
    \end{figure}

\section{\goham: A rotating nebula around a young pre-MS star}

We have seen that it seems impossible to reconcile our knowledge on the
mass and luminosity of Gomez's Hamburger (\gohab) with the assumption
that it is a post-AGB object. The discrepancy is so big that even the
highest corrections to the total luminosity and distance are far from
yielding an acceptable result.

Young stars may also be surrounded by CO-rich rotating disks
\citep[e.g.][]{dutrey07}.  \gohab\ presents properties that, in fact,
seem closer to those of rotating disks around pre-MS A-type stars than
to those of post-AGB nebulae.  The CO line profile (Fig.\ 3) is
relatively symmetric and shows two peaks separated by about 2
\kms. This structure is very similar to that found in AeBe stars
\citep[][etc]{dent05,raman06}, particularly in stars isolated from
interstellar clouds.
However, post-AGB nebulae show in general very wide profiles with
high-velocity wings \citep[e.g.][]{bujetal01}, due to outflowing
gas. Even in the Red Rectangle, the only post-AGB object in which a
keplerian disk has been well identified, and in 89 Her, the best
post-AGB candidate to harbor such a disk after the Red Rectangle, the
CO profiles are different, showing a narrow central peak plus wide
wings. In these cases, the wings are due to expanding outer gas in the
nebulae. We recall that no trace of expanding gas has been found in
\gohab.

We also note that the beautiful optical image of \goham\
(from which its name comes) shows a flat, three-component structure 
remarkably similar to optical images of disks around young stars
\citep[][etc]{grosso03,perrin06,watson07}. The optical image of our source is
however very different from the extended, prolate images of post-AGB
nebula, including that of the Red Rectangle, e.g.\ 
\citet[][]{cohen04}. 

If we assume that the star is a young pre-MS A-type star, some conflict
still persists between the expected values of the stellar mass and
luminosity and those derived from observations, but it is certainly
much less severe.  For a distance of 1 kpc, we have derived a central
stellar mass $\sim$ 8 \ms\ and a total luminosity of about 35 \ls, or
$\sim$ 70 \ls\ for a strong extinction in the line of sight. If we
assume that there is a binary stellar system in the center, a pre-MS
primary with about 4 \ms\ and $\sim$ 10000 K of surface temperature
\citep[value derived from spectroscopic and continuum data
by][]{ruizetal87} should have a luminosity of about 140 \ls, from
comparison with our knowledge on similar young objects \citep[see
compilation of data on other pre-MS sources and evolutionary track
calculations in, e.g.,][]{pietu03,ancker97}.

Measured and expected values of the mass and luminosity are closer if
we assume a shorter distance.  For $D$ = 500 pc, a mass of the primary
of about 2 \ms\ is derived from our CO data (assuming a binary
system). For such a distance, the total luminosity of \gohab\ derived
from its SED is $\sim$ 18 \ls. On the other hand, an A-type pre-MS star
with 2 \ms\ is expected to show luminosities $\sim$ 15 \ls\
\citep{pietu03,ancker97}, compatible with the measured value. We
note that, even in the pre-MS scenario, it is in principle necessary to
assume that there is a binary system in the center of \gohab\ to be
fully in agreement with theoretical evolutionary tracks. Otherwise, we
should have to assume a shorter distance leading to luminosities under
10 \ls, and young stars with such a low luminosity are not expected to
reach a surface temperature of 10000 K.

The stellar mass requirement is somewhat relaxed if we take into
account that the central nebular mass is not negligible. From our
continuum measurement, which is probably due to dust emission
(Sects. 2, 3), we can infer the total mass for the innermost region
(within $\sim$ 1$''$). {From the simple formulation described by
Natta et al.\ (2000, see details on the assumed dust properties in that
paper), we derive a mass of about $0.9 \times \left( \frac{D({\rm
pc})}{500} \right)^2$ \ms. This may particularly affect the mass of
the companion, and our stellar system would be formed by a A0-type star
with $\sim$ 2 \ms\ plus a cooler companion (probably a T Tauri star)
with about 1 \ms, always for a distance of 500 pc.  We note that in the
case of extreme absorption in the line of sight (expected for some
heavy disks around pre-MS stars, Sect.\ 4), which yields a total
luminosity of about 90 \ls\ at 500 pc, only one central star with about
3 \ms\ could explain both the total luminosity and the rotation curve
of \gohab. The location of \gohab\ in the H-R diagram, assuming that it
is a young star and taking into account these uncertainties in the
luminosity, is shown in Fig.\ 4; we also represent well studied AeBe
stars and evolutionary tracks from \citet{ancker98}.

   \begin{figure}
\vspace{-0.2cm}
   \hspace{-0.8cm}
\rotatebox{270}{\resizebox{7.3cm}{!}{ 
\includegraphics{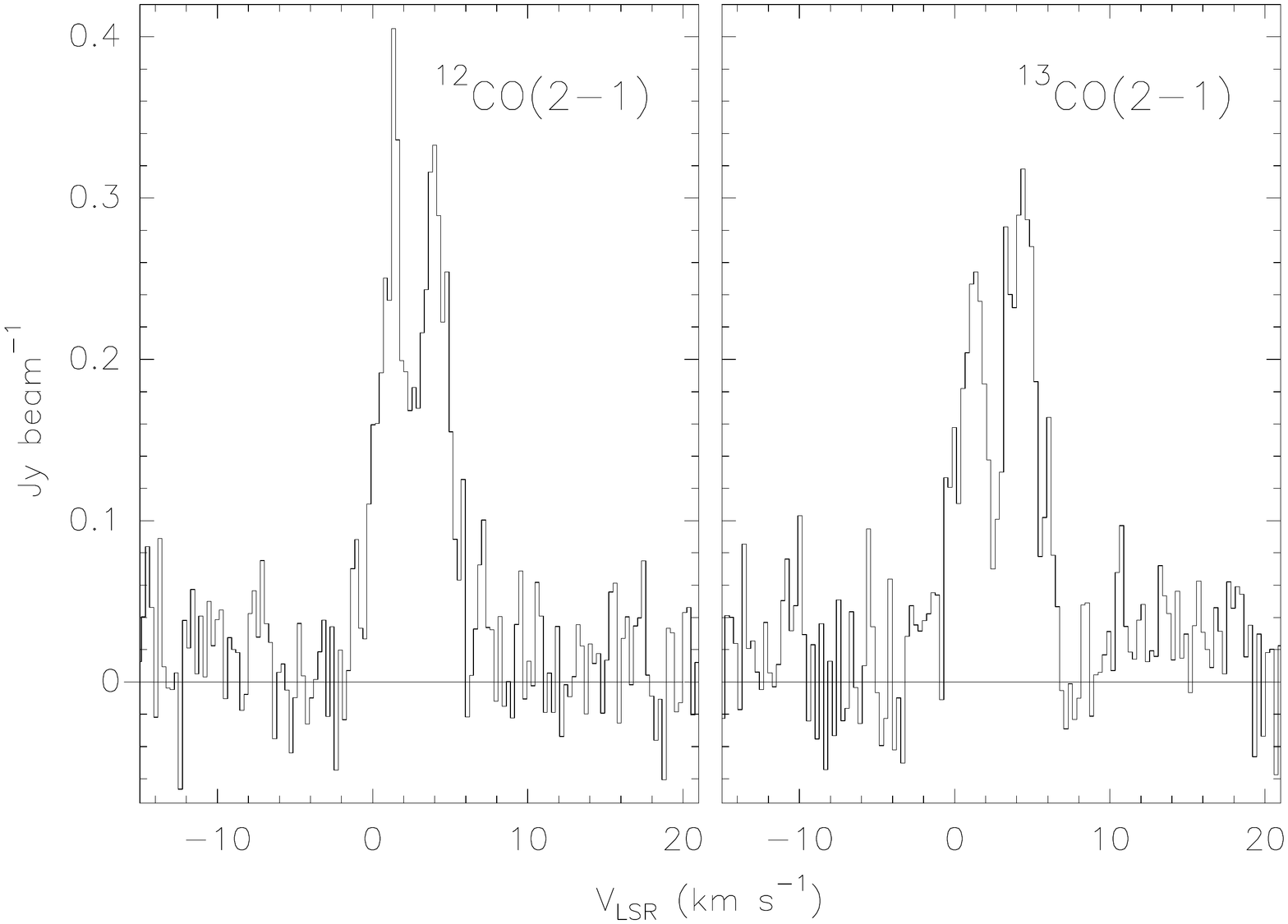}
}}
   \vspace{-0.2cm}
   \caption{\doceCO\ and \treceCO\ \jdu\ line profiles, obtained from integrating in
   our maps over the emitting region.}
              \label{maps}%
    \end{figure}

As noted by \citet{ruizetal87}, Gomez's Hamburger is not associated
with a conspicuous interstellar cloud. But it is within a region of the
sky with a noticeable extinction of about $\sim$ 0.5 mag
\citep[][]{dobashi05}. We also note that many young A-type stars are
not clearly associated with clouds \citep[e.g.][]{dent05}, particularly
in the case of low-luminosity stars, for which the pre-MS evolution is
thought to be relatively slower. The low luminosity, the absence of
conspicuous emission in Balmer lines, and the loose relation with
interstellar clouds strongly suggest that our source is a relatively
evolved star, perhaps close to entering the main sequence.

We do not think that the central star of \gohab\ is a young main
sequence star, similar to the well known debris-disk objects. The
absence of emission in the Balmer lines would support this possibility
\citep[although there is a population of A-type stars assumed to be
very young that do not show such emission features, see 
e.g.][]{the94,manoj06}. But the disks around young MS stars are thought to
disappear relatively quickly (i.e.\ in $\sim$ 10$^7$ yr) and the debris
disks are not massive, yielding a moderate FIR excess and being very
rarely detected in CO emission \citep[e.g.][]{thi01,dent05}.

In summary, we conclude that \gohab\ is very likely a young star
at a distance of about 500 pc, still showing signs of interstellar
material around it. \gohab\ is probably a transitional object between
the pre-MS and MS phases, similar to stars like HD\, 141569, still
surrounded by relatively massive circumstellar disks
\citep[e.g.][]{merinetal04,aug99}. 
In any case, the gas disk orbiting the central star(s) of
\gohab\ is exceptional: it is very massive and extended, and its study
is favored by its observing angle (the disk is seen almost edge-on) and
the lack of contamination due to surrounding interstellar clouds. We
are accordingly convinced that \gohab\ will represent a milestone in
the study of disks rotating around young stars.

   \begin{figure}
\vspace{0.4cm}
   \hspace{0.2cm}
\rotatebox{0}{\resizebox{8cm}{!}{ 
\includegraphics{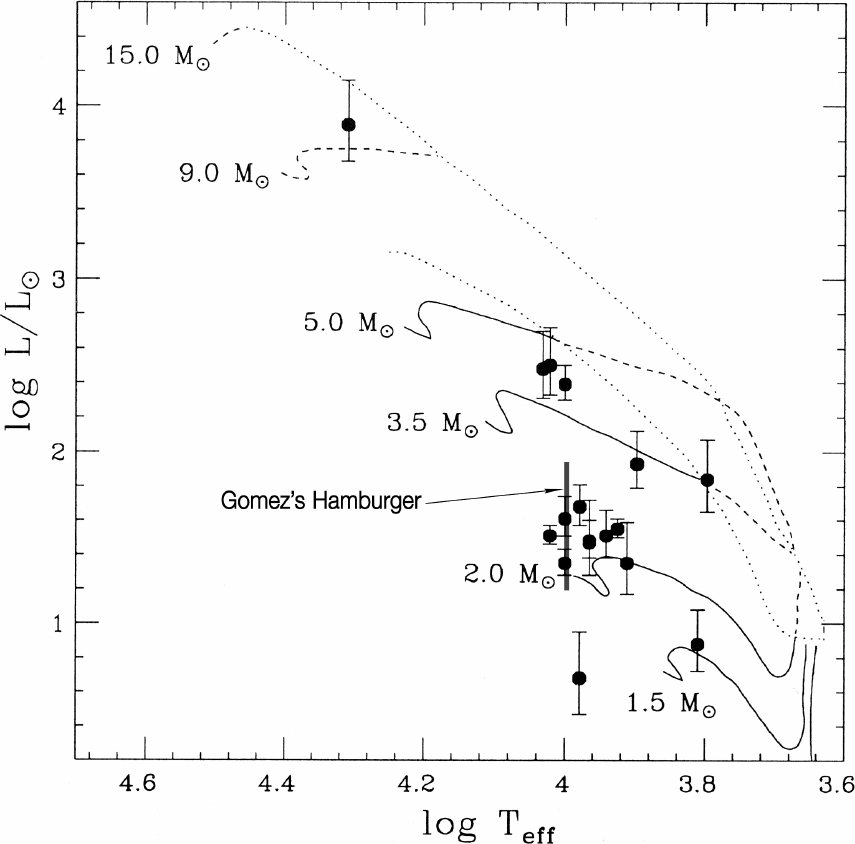}
}}
   \vspace{0.2cm}
   \caption{Location of \goha\ in the H-R diagram, thick vertical line,
   taking into account the uncertainty in the luminosity due to the
   viewing angle effects. We assume a distance of 500 pc, see
   Sect.\ 5; the stellar temperature, $\sim$ 10000 K, is relatively
   well known (see Ruiz et al.\ 1987 and Sect.\ 5). \gohab\ is
   compared with well studied AeBe stars and evolutionary tracks, from
   \citet{ancker98}.}
              \label{maps}%
    \end{figure}



\begin{acknowledgements}
We are grateful to Mario Tafalla and Asunci\'on Fuente for valuable
      discussions at several phases of this work. 
We would also like to thank the anonymous referees of this paper for their
      helpful criticisms. 
      V.B.\ acknowledges support from the \emph{Spanish Ministry of
      Education \& Science} and European FEDER funds, under grants
      AYA2003-7584 and ESP2003-04957.  The Submillimeter Array is a
      joint project between the Smithsonian Astrophysical Observatory
      and the Academia Sinica Institute of Astronomy and Astrophysics
      and is funded by the Smithsonian Institution and the Academia
      Sinica. Some of the data presented in this paper were
      obtained from the Multimission Archive at the Space Telescope
      Science Institute (MAST). STScI is operated by the Association of
      Universities for Research in Astronomy, Inc., under NASA contract
      NAS5-26555. Support for MAST for non-HST data is provided by the
      NASA Office of Space Science via grant NAG5-7584 and by other
      grants and contracts.
\end{acknowledgements}

\bibliographystyle{aa}
\bibliography{protostars}
\end{document}